\documentclass[useAMS,usenatbib,usegraphicx]{mn2e}
\usepackage{amsmath}

\title[Discovery of $\gamma$-ray emission from Pictor A]{Discovery of $\gamma$-ray emission from the Broad Line Radio Galaxy Pictor A}
\author[A. M. Brown and J. Adams]{Anthony M. Brown$^{1}$\thanks{E-mail:
anthony.brown@canterbury.ac.nz} and Jenni Adams$^{1}$\\
$^{1}$Department of Physics and Astronomy, University of Canterbury, Christchurch, 8140, New Zealand}

\begin{document}

\date{Accepted 2011 December 29.  Received 2011 December 22; in original form 2011 November 28}

\pagerange{\pageref{firstpage}--\pageref{lastpage}} \pubyear{2011}

\maketitle

\label{firstpage}

\begin{abstract}
We report the discovery of high-energy $\gamma$-ray emission from the Broad Line Radio Galaxy (BLRG) Pictor A with a significance of $\sim5.8\sigma$ (TS$=33.4$), based on three years of observations with the \textit{Fermi} Large Area Telescope (LAT) detector. The three-year averaged $E>0.2$ GeV $\gamma$-ray spectrum is adequately described by a power-law, with a photon index, $\Gamma$, of $2.93 \pm 0.03$ and a resultant integrated flux of $F_{\gamma}=(5.8\pm0.7) \times 10^{-9}$ ph cm$^{-2}$s$^{-1}$. 

A temporal investigation of the observed $\gamma$-ray flux, which binned the flux into year long intervals, reveals that the flux in the third year was 50\% higher than the three-year average flux. This observation, coupled with the fact that this source was not detected in the first two years of \textit{Fermi}-LAT observations, suggests variability on timescales of a year or less.

Synchrotron Self-Compton modelling of the spectral energy distribution of a prominent hot-spot in Pictor A's western radio lobe is performed. It is found that the models in which the $\gamma$-ray emission originates within the lobes, predicts an X-ray flux larger than that observed. Given that the X-ray emission in the radio lobe hot-spots has been resolved with the current suite of X-ray detectors, we suggest that the $\gamma$-ray emission from Pictor A originates from within its jet, which is in agreement with other $\gamma$-ray loud BLRGs. This suggestion is consistent with the evidence that the $\gamma$-ray flux is variable on timescales of a year or less. 

\end{abstract}


\section{Introduction}

The unified model of Active Galactic Nuclei (AGN) attributes the bright nucleus at the center of some galaxies to a super-massive black hole-accretion disk system surrounded by regions of broad line and narrow line gas clouds. Within this unified model, the difference between radio-loud (RL) and radio-quiet (RQ) AGN is explained by RL AGN possessing a relativistic jet emanating approximately perpendicular from the disk (\citet{ant}; \citet{unified}). 

The early success of the unified AGN model was its ability to explain the observational characteristics of the different RL AGN sub-classes primarily by the size of the angle between the relativistic jet and the observer's line of sight. However, while the unified model works well to first order, there is growing evidence that it is an oversimplification (e.g. \citet{hess2}; \citet{kharb}) and there are still several key fundamental questions that remain to be answered. 

One of these questions is why relativistic jets only appear to be present in 10-20\% of AGN. This question is often included in the broader issue of how accretion and ejecta are linked in AGN (e.g., \citet{bland1}; \citet{bland2}; \citet{sikora}; \citet{mcnam}). To address this issue, the accretion disk-relativistic jet connection needs to be studied across a wide variety of RL AGN types. Broad Line Radio Galaxies (BLRGs) appear to be vital in such a study. With the line of sight orientated at an intermediate angle relative to the relativistic jet, BLRGs exhibit both jet-related and disk-related photon emission processes. BLRGs, therefore allow us to investigate how the processes of accretion (disk) and ejection (jet) relate to each other, and as such, represent a powerful diagnostic tool through which to understand this disk-jet connection\footnote{It should be noted that while the $\gamma$-ray emission processes of radio galaxy jets are thought to be somewhat different to those of blazars, BLRGs still allow us to investigate how the global jet emission is related to the accretion disk processes (\citet{georg}; \citet{ghi}).}.

The successful launch of the \textit{Fermi} $\gamma$-ray Space Telescope affords us an ideal opportunity to investigate the accretion disk-jet connection in AGN. The ability of the \textit{Fermi}-LAT detector to scan the entire $\gamma$-ray sky every three hours allows us to search for $\gamma$-ray emission from AGN unbiased by activity state or AGN sub-class. Recently, the two year LAT AGN catalogue has been released (2LAC), with the 2LAC clean sample consisting of 885 sources (\citet{2fgl}). Of these sources, the vast majority are of the blazar sub-class, however there is a growing number of other AGN sub-classes, including radio galaxies and Narrow Line Seyfert 1 galaxies, detected as $\gamma$-ray sources (\citet{1lac}; \citet{m87}; \citet{nls1}; \citet{nls2}; \citet{kat1275}; \citet{fors}; \citet{fors2}; \citet{brown}). This increase in the number of different AGN sub-classes detected as $\gamma$-ray sources is extremely important for the phenomenological studies required to investigate the connection between accretion and ejection. 

To date, two BLRGs have been detected as $\gamma$-ray sources, 3C 120 and 3C 111 (\citet{blrg}). To investigate the connection between accretion and ejection in BLRGs, \citet{katblrg} were recently able to successfully model the broad-band spectral energy distribution (SED) of both 3C 111 and 3C 120 with a combined model incorporating accretion disk emission, taken from \citet{kor}, and relativistic jet emission, taken from \citet{soldi}. Kataoka et al. also applied this combined disk-jet model to the two-year data set of a sample of 18 X-ray bright BLRGs. While this study did not reveal any new $\gamma$-ray emitting BLRGs, it did find that five of the BLRG sample have flux upper limits that were very close to the flux level predicted by the disk-jet model (\citet{katblrg})\footnote{One of these 5 BLRGs was Pictor A, which hinted at $\gamma$-ray emission with a TS value of 20 from two years of observations.}. Motivated by the proximity of the upper limits to the model prediction, we utlised a larger three-year \textit{Fermi} data set to search for $\gamma$-ray emission from the most promising BLRG candidates. In total four BLRGs sources were studied. While three of these did not show significant $\gamma$-ray emission, our analysis revealed that Pictor A is a high-energy $\gamma$-ray source. 

Pictor A (z=0.035, \citet{pic1}; \citet{pic2}) is an archetypical Fanaroff-Riley II radio galaxy (FRII, \citet{frref}), with the total radio emission dominated by the jet termination region observed as large radio lobes (e.g. \citet{perley}; \citet{tingay1}). From optical spectra, Pictor A is defined as a BLRG (\citet{dann}). Interestingly Pictor A's host galaxy has recently been found to possess a disk-like morphology (\citet{inskip}), indicating that Pictor A is hosted by a spiral galaxy, and thus is one of the rare examples against the AGN jet-ellipical host galaxy paradigm (see \citet{fors2} for more details). Due to its close proximity, Pictor A affords us an excellent opportunity to study not only the disk-jet connection but also the termination shock environment of the radio lobes\footnote{Assuming a $\Lambda$CDM cosmology and $H_{0}=71$ km s$^{-1}$ Mpc$^{-1}$ (\citet{cdm}), Pictor A has a luminosity distance of about 148 Mpc. At this distance 1 arcsecond corresponds to approximately 700 parsecs.}. It is not surprising then that Pictor A has been extensively observed at many wavelengths. 

Radio observations have found Pictor A to possess mildly relativistic jets containing subluminal moving components (\citet{tingay1}). Very Long Baseline Array (VLBA) observations have also shown complex radio structure within Pictor A's radio lobes in the form of compact hot-spots (\citet{tingay2}). These remarkable hot-spots, along with the relativistic jet, have been studied with other observatories, including \textit{Spitzer}, \textit{Hubble Space Telescope}, \textit{XMM} and \textit{Chandra} (e.g. \citet{thomson}; \citet{simkin}; \citet{wilson}; \citet{hardcastle}; \citet{grandi}; \citet{mig}; \citet{tingay2}; \citet{malkan}). These observations have found structured magnetic fields within both the relativistic jet and radio lobe hot-spots. Synchrotron self-Compton (SSC) modelling of these broad-band observations predicts the possibility of $\gamma$-ray emission through inverse-Compton (IC) scattering of the synchrotron photons on the relativistic electrons. However the predicted flux level is somewhat lower than the current $\gamma$-ray upper limits from both the \textit{Fermi} and HESS collaborations (\citet{zhang}; \citet{katblrg}; \citet{hess}). 

While Zhang et al. predicted $\gamma$-ray emission originating from the conditions in the radio lobe hot-spots, the high-energy emission from BLRGs is often attributed to the relativistic jet viewed at intermediate angles (\citet{grandi2}; \citet{sambruna}). The question then arises as to whether the Pictor A $\gamma$-ray emission, that we have discovered, is associated with the radio lobe hot-spots or with the relativistic jet. To address this question we perform SED modelling of a prominent hot-spot in the western radio lobe of Pictor A using archival data at other photon energies. We take two approaches. In our first approach we tune the various parameters available in the SSC model so as to best fit the SED of the radio to optical data and the $\gamma$-ray data. We find that it is not possible to simultaneously fit these photon energy bands as well as the X-ray data with the X-ray flux over-predicted compared to observations. We then take the alternative approach of optimising the fit of the radio to optical data and the X-ray data. We find a good fit to this data but the predicted flux at $\gamma$-ray energies is significantly below the flux that we have observed from Pictor A. This suggests that if the $\gamma$-ray emission is SSC in origin, it is not produced in the radio lobe hot-spot region where the other wavelength photons are observed to originate.
  
The paper outline is as follows: In \textsection 2 we describe the \textit{Fermi}-LAT observations and data analysis routines used in this study. Pictor A's $\gamma$-ray characteristics are given in \textsection 3. In \textsection 4 we present our broad-band spectral energy distribution modelling using archival data at other photon energies, with the discussions on the spectral energy distribution modelling given in \textsection 5. The conclusions are given in \textsection 6. Throughout this paper, a $\Lambda$ cold dark matter ($\Lambda$CDM) cosmology was adopted, with a Hubble constant of H$_{0}=71$ km s$^{-1}$ Mpc$^{-1}$, $\Omega_{m}=0.27$ and $\Omega_{\Lambda}=0.73$ as derived from \textit{Wilkinson Microwave Anisotropy Probe} results (\citet{cdm}).

\section{\textit{Fermi-LAT} observations and Data Analysis}
The LAT detector aboard \textit{Fermi}, described in detail by \citet{lat}, is a pair-conversion telescope, sensitive to a photon energy range from below 20 MeV to above 300 GeV. With a large field of view, $ \simeq 2.4 $ sr, improved angular resolution, $\sim0.8$\ensuremath{^{\circ}} at 1 GeV\footnote{Below 10 GeV photon energy, the 68\% containment angle of the photon direction is approximately given by $\theta \simeq 0.8$\ensuremath{^{\circ}}($E_\gamma/$GeV)$^{-0.8}$, with the 95\% containment angle being less than 1.6 times the angle for 68\% containment.}, and large effective area, $\sim 8000$ cm$^{2}$ on axis at 10 GeV, \textit{Fermi}-LAT provides an order of magnitude improvement in performance compared to its \textit{EGRET} predecessor. 

Since 2008 August 4, the vast majority of data taken by \textit{Fermi} has been performed in \textit{all-sky-survey} mode, whereby the \textit{Fermi}-LAT detector points away from the Earth and rocks north and south of its orbital plane. This rocking motion, coupled with \textit{Fermi}-LAT's large effective area, allows \textit{Fermi} to scan the entire $\gamma$-ray sky every two orbits, or approximately every three hours (\citet{ritz}). 

The data used in this analysis comprises of all \textit{all-sky-survey} observations taken during the first $\sim$three years of \textit{Fermi} operation, between 2008 August 4 and 2011 August 8, equating to a mission elapse time (MET) interval of 239557417 to 334486765. In accordance with the PASS 7 criteria, a zenith cut of 100\ensuremath{^{\circ}} along with a rock-angle cut of 52\ensuremath{^{\circ}} was applied to the data to remove any cosmic ray induced secondary $\gamma$-rays from the Earth's atmosphere. All `source' class events\footnote{`Source' class events equates to EVENT\_CLASS$=$2 in the PASS 7 data set. Source class events optimise the $\gamma$-ray detection efficiency as a function of residual charged-particle contamination and is best suited for localised sources as opposed to transient or diffuse sources.}  within a 6\ensuremath{^{\circ}} radius of interest (ROI) were considered in the $0.2 < E_{\gamma} < 300$ GeV energy range, with the lower energy cut being conservative to reduce the systematic errors.

Throughout this analysis, the most recent \textit{Fermi}-LAT Science Tools, version v9r23p1, were used in conjunction with instrument response functions P7SOURCE\_V6. Furthermore, the most recent galactic, gal\_2yearp7v6\_v0.fits, and extragalactic, iso\_p7v6source.txt, diffuse models were utilised during spectral fitting and upper limit calculations.

\section{$\gamma$-ray characteristics}

To investigate the $\gamma$-ray properties of Pictor A, we utilized the unbinned maximum likelihood estimator of the \textit{Fermi}-LAT Science Tools' GTLIKE routine. GTLIKE allows us to fit the data with a series of both point and diffuse sources of $\gamma$-rays. The model used to calculate the likelihood of $\gamma$-ray emission from Pictor A was a combination of the most recent galactic and extragalactic diffuse models, as given in \textsection 2, and all point sources within an 8\ensuremath{^{\circ}} radius region of interest (ROI) centered on Pictor A. Each point source was modelled with a power-law spectrum of the form $dN/dE = $ A$ \times E^{-\Gamma}$ with both parameters, photon index, $\Gamma$, and normalisation, A, free to vary. The co-ordinates for each point source within the ROI were taken from the 2nd \textit{Fermi}-LAT Catalogue (2FGL; \citet{2fgl}). In addition, a 10\ensuremath{^{\circ}} radius sky map of the three year data set was used to search for any other sources in the ROI that were not present in the 2FGL. 

As a further check, the spatial distribution of the test statistic\footnote{The test statistic, TS, is defined as twice the difference between the log-likelihood of two different models, $2[\text{log} L - \text{log} L_{0}]$, where $L$ and $L_{0}$ are defined as the likelihood when the source is included or not respectively (\citet{mattox}).} values obtained from the GTLIKE fit was also utilised to identify any other possible sources in the ROI that were not present in the 2FGL. The TS distribution indicated the presence of two TS peaks within 1\ensuremath{^{\circ}} of Pictor A. One of these peaks has a TS value of $76.6$ and is spatially coincident with the blazar BZQ J0515$-$4556. Indeed, BZQ J0515$-$4556 is present in the 2FGL and therefore was already included in the model file. 

The other TS peak is separated from Pictor A by $\sim0.9$\ensuremath{^{\circ}}, located at the position ($\alpha_{J2000}$, $\delta_{J2000}=81.3$\ensuremath{^{\circ}}, $-45.9$\ensuremath{^{\circ}}). With a TS value of $13.8$, or $\sim3.7\sigma$, this TS peak suggests the presence of another $\gamma$-ray source which is not in either the 1FGL or 2FGL catalogues. However given that the TS value is below the detection threshold of 25, this emission is not yet significant enough to be formally classified as a $\gamma$-ray source. 

Another \textit{Fermi}-LAT Science Tool, GTFINDSRC, was applied to the 8\ensuremath{^{\circ}} ROI around Pictor A to localise the origin of the $\gamma$-ray emission. Using the same combined diffuse and point source model that was applied during the GTLIKE routine, the $\gamma$-ray emission that is associated with Pictor A is located at ($\alpha_{J2000}$, $\delta_{J2000}=80.17$\ensuremath{^{\circ}}, $-45.61$\ensuremath{^{\circ}}), with a 95\% error radius of 0.08\ensuremath{^{\circ}}. The GTFINDSRC routine was also able to identify and localise the emission from both BZQ J0515$-$4556 and the other TS `hot-spot' as two $\gamma$-ray sources separate from Pictor A.

Once satisfied that all point sources within an 8\ensuremath{^{\circ}} ROI have been accounted for, we applied GTLIKE to the three-year data set. With the combined diffuse and 8\ensuremath{^{\circ}} ROI point source model, we obtain the following best-fit power-law function for Pictor A:

\begin{equation}
 \dfrac{dN}{dE}= (4.23 \pm 0.44) \times 10^{-10} (\dfrac{E}{100\text{ MeV}})^{-2.93\pm0.03} \nonumber
\end{equation}

\begin{equation}
 \text{ photons cm}^{-2} \text{ s}^{-1} \text{ MeV}^{-1}
\end{equation}
which equates to an integrated flux, in the $0.2 < E_{\gamma} < 300$ GeV energy range, of
\begin{equation}
  F_{E>0.2\text{ GeV}} = (5.8 \pm 0.7) \times 10^{-9}  \text{ photons cm}^{-2} \text{ s}^{-1}
\end{equation}
only taking statistical errors into account. Primarily governed by the uncertainty in the effective area, the systematic uncertainty of the integrated flux is energy dependent and is currently estimated as 10\% at 100 MeV, down to 5\% at 560 MeV and back to 10\% for 10 GeV photons (\citet{2fgl2}).

From the power-law fit, the predicted $\gamma$-ray count was 337, with a test statistic of $TS=33.4$, corresponding to a $\sim 5.8\sigma$ detection. Compared to other $\gamma$-ray BLRGs, 3C 120 and 3C 111, Pictor A's $\gamma$-ray flux is a factor of six lower, and the spectral index is comparable, with a spectral index of $-2.93\pm0.03$ for Pictor A compared to $-3.0\pm0.3$ for 3C 120 and $-2.7\pm0.2$ for 3C 111. On the other hand, the $\gamma$-ray luminosity for Pictor A is comparable to both 3C 120 and 3C 111, log($L_{\gamma}$)$=43.1$ ergs s$^{-1}$ compared to 43.6 ergs s$^{-1}$ and 43.9 ergs s$^{-1}$ respectively (\citet{katblrg}).

\begin{figure}
 \centering
 \includegraphics[width=95mm]{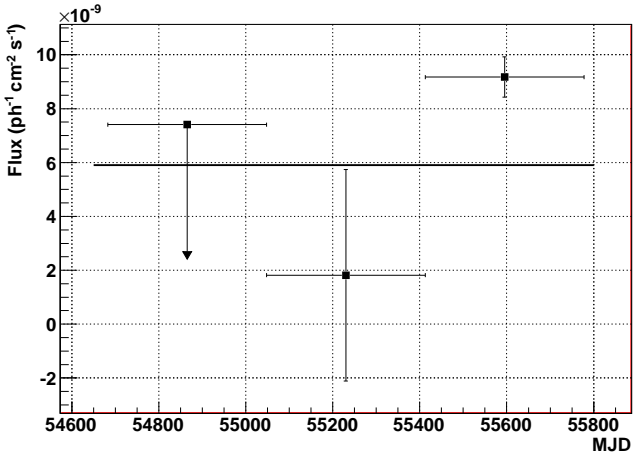}
 \caption{Light curve of the $0.2 < E_{\gamma} < 300$ GeV $\gamma$-ray flux from Pictor A during the first three years of \textit{Fermi}-LAT observations. The data is binned in yearly intervals, with the fluxes being plotted when the TS value is greater than 10 ($\sim3\sigma$), otherwise the 95\% confidence upper limit is given. The bold horizontal line is the integrated flux given in Equation 2, derived from the three-year power-law fit. The difference in flux levels between the third annual bin and the three-year averaged flux, as well as between the second and third year, suggests variability on timescales of a year or less, though it should be stressed that further observations are required to confirm this.}
\label{lc}
\end{figure}

To investigate the possibility of temporal variability in Pictor A's $\gamma$-ray flux, the three-year data set was binned into one year periods and analysed separately with GTLIKE. As with the spectral analysis, only $0.2 < E_{\gamma} < 300$ GeV events were considered, with the appropriate quality cuts applied, and the same combined diffuse and 8\ensuremath{^{\circ}} ROI point source model was used. Only the fluxes of temporal bins with TS $>10$ ($\sim3\sigma$) were plotted, with the 95\% confidence upper limit being calculated for the remaining temporal bin. The resultant light curve can be seen in Figure \ref{lc}, with the bold horizontal line indicating the three-year averaged $0.2 < E_{\gamma} < 300$ GeV flux level.

A closer inspection of the individual photon energies reveals that the highest photon energy detected from Pictor A during this three year data set has an energy of 109.8 GeV. Additionally, photons with energies of 99.7 and 91.3 GeV were also detected, both within 0.8\ensuremath{^{\circ}} of Pictor A. The spatial distribution of these highest energy photon events was compared to the sky map to confirm that none were spatially coincident with either BZQ J0515$-$4556 or the $13.8$ TS peak. Taking the approach outlined in \citet{neronov}, the chance probability of the second highest photon being clustered within 1\ensuremath{^{\circ}} of the highest photon is $\approx 0.107$. That is to say, the chance probability that these two photon events are not clustered background events is  $\approx 89.3$\%. This value is $\sim 2 \sigma$ and therefore the observed `clustering' of $\ga 90$ GeV photons around Pictor A is not statistically significant. It is worth noting that a GTLIKE likelihood analysis for all very high energy (VHE) $\gamma$-ray events, $90 < E_{\gamma} < 300$ GeV, from Pictor A, utilising the same combined diffuse and 8\ensuremath{^{\circ}} ROI point source model, finds a comparable statistical significance to Pictor A being a VHE $\gamma$-ray source, with a TS value of 4.99 or $\sim 2.2 \sigma$.

\section{Spectral energy distribution modelling}

We now turn to the question of whether the observed $\gamma$-ray emission is associated with the hot-spots in the terminal shocks of the radio lobes or with the relativistic jet. \textit{Fermi}-LAT has resolved $\gamma$-ray emission from the radio lobes of the low-power radio galaxy, Centaurus A (\citet{cena}). However, beyond Centaurus A, further evidence for $\gamma$-ray emission from AGN radio lobes is scarce.

As mentioned in \textsection 1, leptonic models which predict $\gamma$-ray emission from the radio lobe hot-spots attribute the observed radio-UV emission to synchrotron radiation, primarily due to the observation of polarised radiation from the hot-spots at these wavelengths (eg. \citet{roeser}; \citet{harris}). Given the synchrotron origin of this radiation, the shape of the SED at radio to UV wavelengths places strong restrictions on the broken power-law energy distribution of the electron population within the hot-spot. The $\gamma$-ray emission, in such a model, is a result of the IC process acting on a percentage of the synchrotron photons.

Due to the close proximity of Pictor A, radio lobe hot-spot features have been individually resolved at X-ray, optical, infrared and radio wavelengths\footnote{The angular resolution of \textit{Fermi} does not allow us to resolve Pictor A's radio lobes and nucleus regions separately.}. These observations have discovered several bright knots within Pictor A's radio lobe, with the Western hot-spot (WHS) being the most prominent and thus most extensively studied. SSC modelling of the WHS spectral energy distribution (SED) has already been performed by \citet{zhang}, who predicted a GeV $\gamma$-ray flux level on the order of $\sim10^{-11.5}$ ergs cm$^{-2} \text{ s}^{-1}$. This prediction is somewhat higher than the three-year averaged flux that we have determined in the previous section. 

Given there are a number of tuneable parameters available in the SSC model we performed SSC modelling to determine the consistancy of the observed $\gamma$-ray flux being produced through IC in the radio lobe hot-spots. The SED was constructed using historical radio, infra-red, optical and X-ray observations that have reported flux measurements from the WHS alone (\citet{meisen}; \citet{tingay2}; \citet{mig}). We also included the VHE $\gamma$-ray upper limit set by the HESS collaboration (\citet{hess}). The SSC model assumed a spherical emission region of radius $R$, moving with Doppler factor $\delta$ and filled with a random magnetic field of strength $B$ and an electron population described by a broken power-law (for more details, see \citet{ssc}). The radius of the emission region was fixed to the physical size of the WHS, $R=209$ pc, as observed with the \textit{Hubble Space Telescope} (\citet{thomson}) and the Doppler factor, $\delta$, was fixed to 1 thus removing any doopler boosting effects. 

\begin{figure}
 \centering
 \includegraphics[width=60mm,angle=270]{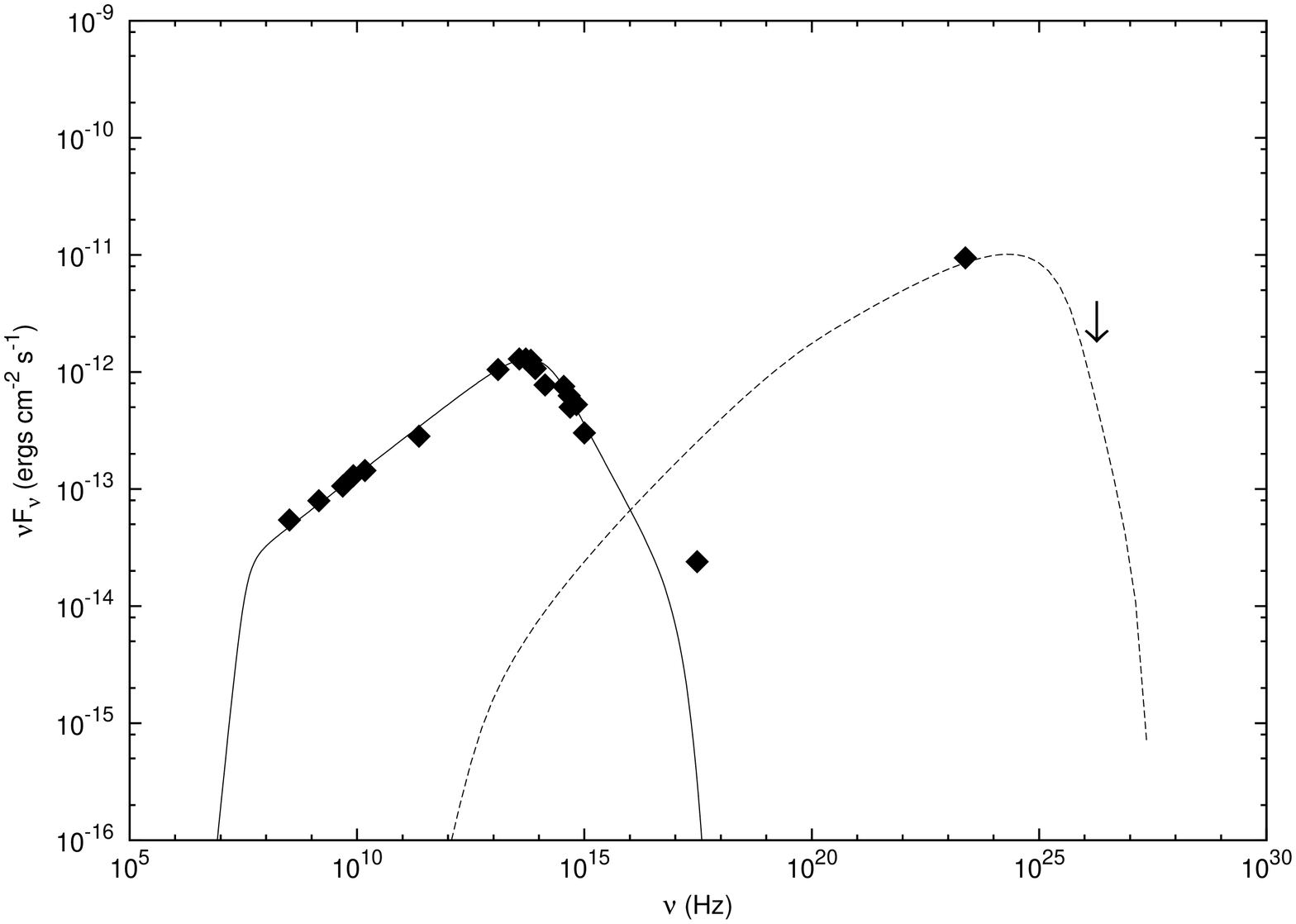}
 \caption{SSC modelling of the SED of Pictor A's WHS from radio to $\gamma$-ray, plotted as $\nu F_{\nu}$. The optical data is taken from \citet{meisen}, the radio and infrared data is taken from \citet{tingay2}, the X-ray data is taken from \citet{mig}, the VHE $\gamma$-ray upper limit, marked by an arrow, is taken from \citet{hess} and the \textit{Fermi} data point is report here. The solid line is the synchrotron spectrum and the dashed line is the inverse Compton spectrum. The SSC fit has been optimised to describe the radio to optical and the $\gamma$-ray flux.}
\label{hot}
\end{figure}

In our first approach, the SSC fit to the WHS SED, shown in Figure \ref{hot}, was originally optimised to account for the synchrotron radiation distribution, whilst accurately modelling our \textit{Fermi}-LAT observations. The WHS SED was found to be best described by an electron distribution of $E_{min}=10^8$ eV, $E_{break}=4\times10^{11}$ eV and $E_{max}=10^{13}$ eV, with broken power-law indices of $n_1=2.4$ and $n_2=4.4$ below and above the break energy respectively. To reproduce the relative synchrotron and inverse Compton peak flux levels, a magnetic field of $4\times10^{-5}$ Gauss and an electron energy density of $2.5\times10^{-6}$ ergs cm$^{-3}$ was required. The resultant SSC fit can be seen in Figure \ref{hot} with the solid line representing the synchrotron spectrum and the dashed line representing the IC spectrum. 

As can be seen in Figure \ref{hot}, while the best fit SSC model adequately describes the radio-optical and \textit{Fermi} fluxes, it over-predicts the inverse Compton flux at X-ray energies by an order of magnitude. It is worth noting at this point, that while smaller, this discrepancy is also present if the entire Western radio lobe's X-ray flux is included in the SED (\cite{mig}). 

\begin{figure}
 \centering
 \includegraphics[width=60mm,angle=270]{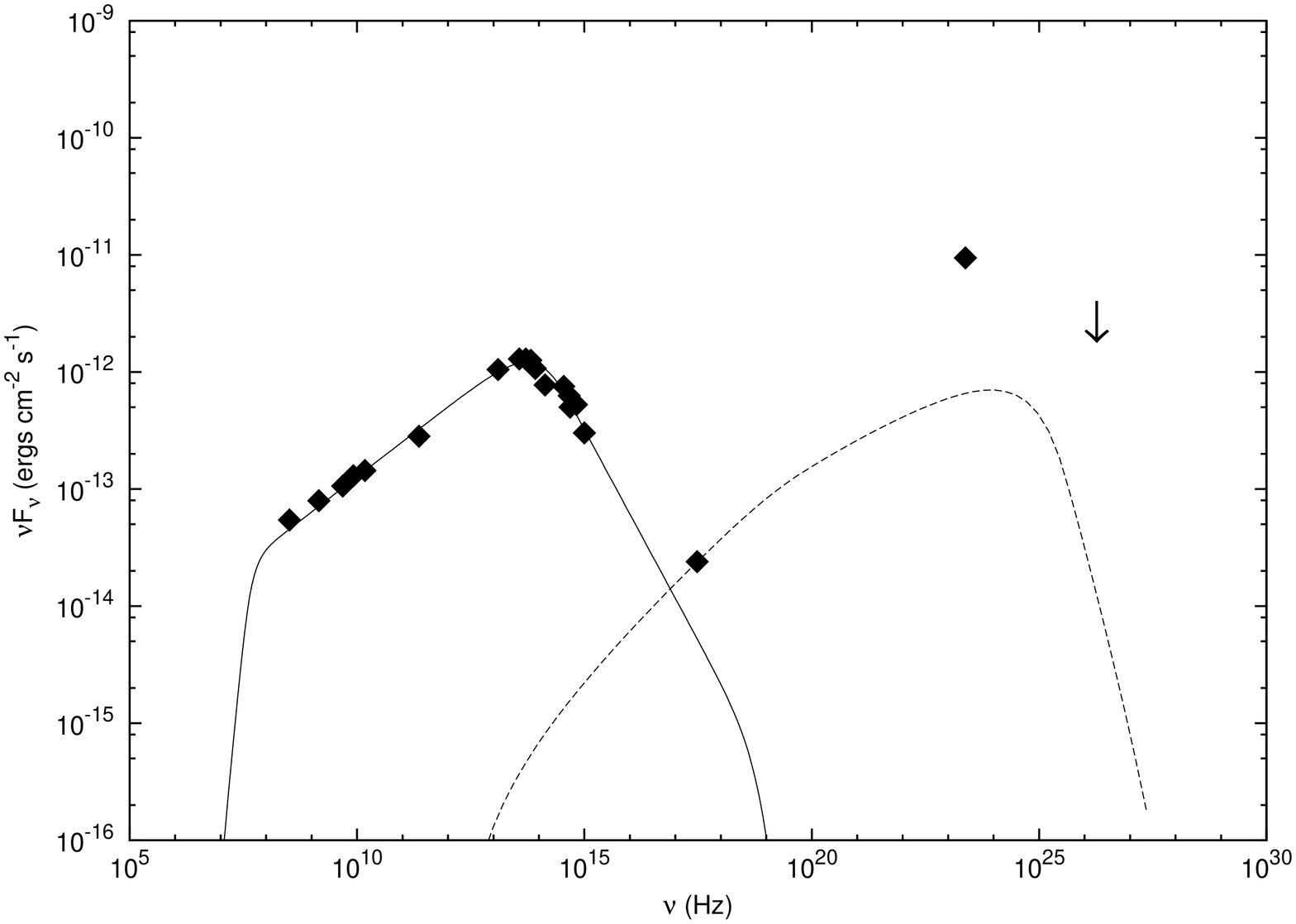}
 \caption{SSC modelling of the SED of Pictor A's WHS from radio to $\gamma$-ray, plotted as $\nu F_{\nu}$. The optical data is taken from \citet{meisen}, the radio and infrared data is taken from \citet{tingay2}, the X-ray data is taken from \citet{mig}, the VHE $\gamma$-ray upper limit, marked by an arrow, is taken from \citet{hess} and the \textit{Fermi} data point is report here. The solid line is the synchrotron spectrum and the dashed line is the inverse Compton spectrum. The SSC fit has been optimised to describe the radio to X-ray flux.}
\label{hot2}
\end{figure}

To address the over-prediction of the X-ray flux, both the magnetic field strength and electron population's energy distribution were varied, optimising the fit for the synchrotron radiation distribution, whilst accurately modelling the X-ray observations. The best fit for the X-ray data was described an electron distribution of $E_{min}=10^8$ eV, $E_{break}=2\times10^{11}$ eV and $E_{max}=4\times10^{13}$ eV, with broken power-law indices of $n_1=2.4$ and $n_2=4.4$, a magnetic field of $1.5\times10^{-4}$ Gauss and an electron energy density of $2.5\times10^{-7}$ ergs cm$^{-3}$. It should be noted that a slight change in the break and maximum energies of the electron population were required so as to not over predict the flux of the high energy synchrotron emission. The resultant `X-ray' fit can be seen in Figure \ref{hot2}. The fit parameters for Figure \ref{hot} and Figure \ref{hot2} are shown in Table \ref{comparison}, along with the fit parameters for Zhang et al.'s modelling of Pictor A's WHS and the \textit{Fermi}-LAT Collaboration's modelling of Cen A's core and radio lobes.

\begin{table*}
 \begin{minipage}{150mm}
   \caption{Summary of fit parameters. The Cen A core fit parameters are taken from \citet{cena} while the Cen A radio lobe fit parameters are taken from \citet{cena2}. It should be noted that the Cen A radio lobe fit considers 4 separate emission regions, 2 for each radio lobe; we quote the fit parameters for the total northern lobe, since the particle energetics for the two emission regions in the north lobe are the same. The Zhang et al. Pic A fit parameters are taken from the SSC fit parameters in \citet{zhang}. The energy limits of the electron population ($E_{min}$, $E_{break}$ \& $E_{max}$) from our fits have been converted to Lorentz factors ($\gamma_{min}$, $\gamma_{break}$ \& $\gamma_{max}$) for comparison to the Cen A fits.}
   \begin{center}
     \begin{tabular}{llllll} \hline \hline
               & Pic A (X-ray fit) & Pic A (\textit{Fermi} fit) & Pic A (Zhang et al.)  & Cen A$_{core}$ & Cen A$_{lobe}$  \\ \hline
   Magnetic field (G)    & $1.5\times 10^{-4}$  & $4\times 10^{-5}$ & $4.4\times 10^{-5}$   & 6.2 & $8.9\times 10^{-4}$ \\ 
   Doppler factor ($\delta$) & 1 & 1  & 1 & 1 & $-$ \\
   volume (pc$^3$)& $2.14 \times 10^{-7}$ & $2.45 \times 10^{-8}$ & $4.1 \times 10^{-7}$ & $3.8 \times 10^{-9}$ & $3.6\times10^{15}$\\ 
   $\Gamma_1$   & 2.4 & 2.4 & 2.38  & 1.8 & 2.1 \\
   $\Gamma_2$  & 4.4 & 4.4 & 3.9 & 4.3 & 3.0 \\ 
   $\gamma_{min}$   & 196 & 196 & $-$ & 300 & 1 \\
   $\gamma_{break}$   & $7.8\times 10^{5}$ & $3.9\times 10^{5}$ & $-$ & 800 & $3.6\times 10^{4}$  \\
   $\gamma_{max}$   & $2\times 10^{7}$ & $7.8\times 10^{7}$ & $-$ & $1\times 10^{8}$ & $3.3\times 10^{5}$  \\\hline
    \end{tabular}
  \end{center}
  \label{comparison}
\end{minipage}
\end{table*}

\section{Discussion}

In both Figure \ref{hot} and Figure \ref{hot2}, the radio to optical emission can be seen to be accurately described by the process of synchrotron radiation. It is not, however, possible to simultaneously describe the X-ray and $\gamma$-ray fluxes by the IC component of the one-zone SSC model. Therefore, our SED modeling provides support for the synchrotron origin of the low energy emission of the WHS of Pictor A. However, assuming that the SSC model is the dominant mechanism for the observed X-ray flux from the WHS, then our SED modelling suggests that Pictor A's observed $\gamma$-ray emission is either produced through processes other than SSC, or that the $\gamma$-ray originates from either Pictor A's relativistic jet or central core region. 

Other possible mechanisms through which the $\gamma$-rays can be emitted include the external-Compton (EC) model or hadronic models invoking photon-proton or proton-proton interactions. Similiar to the SSC model, the EC model simply has an external photon field, such as the Cosmic Microwave Background (CMB), as the dominant photon field on which the inverse-Compton process acts. It is worth noting that  Migliori et al. attributed the X-ray emission from the WHS to the EC of the CMB (\citet{mig}), however more recently Zhang et al. found the EC contribution from the CMB to be negligible (\citet{zhang}). While we have not considered EC during our SSC modelling of Pictor A's WHS, doing so would simply allow us to relax the energetic requirements of the electron population in the hot-spot. 

In depth hadronic interpretation of the \textit{Fermi}-LAT observations is outside the scope of this paper. Nonetheless, we do note that  neutrino telescopes, such as ANTARES and IceCube, have the potential to constrain the hadronic contribution to Pictor A's observed $\gamma$-ray flux (e.g. \citet{becker}); \citet{icecube}; \citet{icecube2}; \citet{antares}). 

However, it seems more likely that the reason why our SED modelling was unable to marry the WHS observations with the observed $\gamma$-ray flux is that the $\gamma$-ray emission is of jet origin. As mentioned in \textsection 1, $\gamma$-ray emission from BLRGs is traditionally interpreted as originating from a relativistic jet viewed at intermediate angles from the line of sight (e.g. \citet{grandi2}; \citet{sambruna}; \citet{katblrg}). Flux variability is a fundamental property of astrophysical jets. Indeed, evidence for variability in Pictor A's jet has been observed with \textit{Chandra} which has observed the X-ray flux from Pictor A's western jet to vary on timescales of years (\citet{marshall}). Evidence for variability in Pictor A's $\gamma$-ray flux can be seen in Figure \ref{lc}, which hints at $\gamma$-ray flux variability on timescales of a year or less. This evidence is primarily seen in the third year of observations, with a $\sim4\sigma$ detection of Pictor A and a flux level that is 50\% larger than the 3 year average\footnote{It is also worth noting that the flux in the third year is approximately 30\% larger than the 95\% flux upper limit set in the first year of observations. Furthermore, the 3$\sigma$ flux in the second year, is over five times smaller than the flux observed in the third year of observations.}.Taking into consideration statistical errors, the difference between the third year and three-year averaged flux levels is significant at the $\sim4\sigma$ level. However, this is just short of the $5\sigma$ level, and as such, further \textit{Fermi}-LAT observations are need to confirm this observational characteristic. 

At this point, we briefly turn our attention to other possible locations for the $\gamma$-ray emission, namely the core region of Pictor A within the immediate vicinity of the super-massive black hole (SMBH). Magnetospheric emission models of AGN core regions have been successful invoked to explain the $\gamma$-ray emission from other radio galaxies such as M87 (e.g. \citet{rieger}). However, as highlighted in \citet{brown}, for these models $L_{IC} \propto M_{BH}$ is expected, where $L_{IC}$ is the IC luminosity, which in this case is the $\gamma$-ray luminosity and $M_{BH}$ is the SMBH mass. Given that Pictor A's SMBH mass is an order of magnitude larger then Cen A's, whilst the $\gamma$-ray flux is two orders of magnitude smaller, we find it unlikely that a magnetospheric emission model significantly contributes to the observed $\gamma$-ray emission and suggest that the observed $\gamma$-ray emission from Pictor A is indeed of jet origin.

Therefore, given that \textit{Chandra} has resolved X-ray emission from the WHS, the SSC fit of Figure \ref{hot2} indicates that the observed $\gamma$-ray flux cannot be attributed to an SSC processes within the hot-spots of Pictor A's radio lobes. This conclusion, coupled with the evidence of variability observed in Figure \ref{lc}, suggest that the $\gamma$-ray emission from Pictor A originates from within its jet, in agreement with the SED modelling of other $\gamma$-ray loud BLRGs.

\section{Conclusions}

We have reported the discovery of Pictor A as a $\gamma$-ray source with a significance of $5.8 \sigma$, utilising the first three years of observations with the \textit{Fermi}-LAT detector. These observations have found the $E> 0.2$ GeV spectrum to be well described by a power-law with $\Gamma = -2.93\pm0.03$, which is comparable to the other \textit{Fermi}-LAT detected BLRGs, 3C 120 and 3C 111. Pictor A's three-year averaged $0.2 < E_{\gamma} < 300$ GeV flux is $(5.8 \pm 0.7) \times 10^{-9} \text{ photons cm}^{-2} \text{s}^{-1}$, which equates to a $\gamma$-ray luminosity of log($L_{\gamma}$)$=43.1$ ergs s$^{-1}$. Pictor A's  $\gamma$-ray flux was thus found to be approximately six times smaller than 3C 120 and 3C 111, but due to its close proximity, a luminosity that is comparable with other $\gamma$-ray loud BLRGs.

Pictor A's year-binned light curve suggested variability in the $\gamma$-ray flux, with the third year flux level being 50\% larger than the three year average. However, further observations are needed to confirm this observational property. 

SSC modelling of the WHS SED found that it was only possible to account for the $\gamma$-ray emission with an SSC model at the expense of greatly over-predicting the flux at X-ray energies. Given that the X-ray emission in the WHS has been resolved with the current suite of X-ray detectors, we suggest that the $\gamma$-ray emission from Pictor A originates from within its jet, in agreement with other BLRGs detected with \textit{Fermi}. This suggestion agrees well with the evidence that the $\gamma$-ray flux is variable on timescales of a year or less. 

The highest energy photon detected during this 3 year period was 109.8 GeV, with several $>$90 GeV photons also being observed. A likelihood analysis of the $90< E_{\gamma} <300$ GeV $\gamma$-ray events does not find any significant VHE emission. While the clustering of the highest energy events does not appear to be significant, it raises the interesting possibility of VHE $\gamma$-ray emission from Pictor A.

\section*{Acknowledgments}

This work is supported by the Marsden Fund Council from New Zealand Government funding, administered by the Royal Society of New Zealand. The authors would like to thank the referee, Luigi Foschini, for his useful suggestions that helped improve this manuscript. The authors would also like to thank Gianfranco Brunetti and Steven Tingay for providing the radio-optical SED data used in Figures 2 \& 3. This work has made use of public \textit{Fermi} data obtained from the High Energy Astrophysics Science Archive Research Center (HEASARC), provided by NASA's Goddard Space Flight Center. This work has also made use of the NASA/IPAC Extragalactic Database (NED), which is operated by the Jet Propulsion Laboratory, Caltech, under contact with the National Aeronautics and Space Administration, as well as SAO/NASA's Astrophysics Data System (ADS), hosted by the High Energy Astrophysics Division at the Harvard-Smithsonian Center for Astrophysics.

\end{document}